\documentclass[aps,amsmath,citeautoscript,longbibliography,nofootinbib]{revtex4-2}
\usepackage[utf8]{inputenc}
\usepackage{bm}
\usepackage{tensor}
\usepackage[italicdiff]{physics}
\usepackage{amsmath,mathrsfs,amsfonts}
\usepackage{graphicx,afterpage}
\usepackage{subcaption}
\usepackage{amsmath,latexsym}
\usepackage{color,soul}
\usepackage{float}
\usepackage[section]{placeins}
\usepackage{silence}
\usepackage[hidelinks]{hyperref}
\usepackage{lipsum}
 \newcommand\yesnumber{\addtocounter{equation}{1}\tag{\theequation}}
\DeclareMathOperator{\diag}{Diag}
\begin{document}
\title{Exact Analytical Taub-NUT like solution in $f(T)$ gravity}
\author{Joshua G. Fenwick}
\email{joshua.fenwick@usask.ca}
\author{Masoud Ghezelbash}
\email{masoud.ghezelbash@usask.ca}
\affiliation{Department of Physics and Engineering Physics, University of Saskatchewan, Saskatoon SK S7N 5E2, Canada}
\date{\today}
\begin{abstract}
 We construct exact analytical Taub-NUT solutions in the context of $f(T)$ gravity. We study the physical properties of the solutions, and compare them with those of the Taub-NUT solution in Einstein gravity.
\end{abstract}
\maketitle
\section{Introduction}
The Modified Theories of Gravity (MTG), which have been known for a long time, are theories to address the issues such as dark energy, since the accelerating expansion of the universe was discovered. These theories are of great importance in cosmology, to handle the problems, for which general relativity does not provide any solutions. Moreover, in the appropriate limits, these theories should produce the same results as general relativity. The first modified theory of gravity is $f(R)$ theory, in which the Ricci scalar is replaced by a general function $f(R)$, in the Einstein-Hilbert action \cite{24,25,26,27,28,29,30,31,32,34,35,36,37,38,39,40,41}. The second such modified theory is to use the curvature-free connection, instead of the usual torsion-free Levi-Civita connection in general relativity. The simplest possibility is to use the torsion scalar $T$ , instead of Ricci scalar in the action of general relativity. The theory is called Teleparallel Equivalent of General Relativity (TEGR). We can consider a more general theory, where the action is a function of torsion scalar $f(T)$, which leads to a new class of MTG \cite{Spherical,43,44,45,46,47,48,Biancci_identity,50,51}. The third modified theory is to use the non-metricity, instead of curvature or torsion to describe the gravity. The simplest possibility is to use the non-metricity scalar Q, instead of Ricci or Torsion scalars in the action of general relativity. The theory is called Symmetric Tele-parallel Equivalent of General Relativity (STEGR). Quite interestingly, TEGR and STEGR are equivalent to general relativity. However considering a more general theory, where the action is a function of non-metricity scalar $f(Q)$, leads to a new class of MTG \cite{52,53,54,55,56}. Moreover, we can consider other more general theories of gravity, where the action is a function of more than one of the $R$, $T$ and $Q$. As an example, $f(R,T)$ theories were considered in cosmological model building and observational constraint \cite{57,58,59,60,61,62,63,64,65,66}. 
\textcolor{black}{
In particular, the $f(T)$ theories of gravity have been used extensively to describe the behaviour of the universe at cosmic scales, which involve more advanced geometries beyond the traditional Riemannian geometry.  As few examples in $f(T)$ gravity, we should mention articles about finding new cosmological solutions \cite{6615},  reconstruction, thermodynamics and stability of the $\Lambda$CDMmodel \cite{6616}, equation of state for dark energy \cite{6617}, cosmological perturbations \cite{6619} and local Lorentz invariance \cite{6620}. Moreover we should mention the following article in $f(T)$ gravity, on cosmography \cite{6621}, energy conditions bounds \cite{6622}, matter bounce cosmology \cite{6623},  stability analysis of anisotropic Bianchi type-I cosmological model in teleparallel 
gravity \cite{6626}, anisotropic universe models \cite{6625} and exploring the late-time cosmic acceleration through linear  cosmological model using observational data \cite{6628}.  Some other articles in $f(T)$ gravity are focused on exploring cosmological evolution and constraints \cite{6630}, resolving FLRW cosmology through effective equations of state \cite{6631},  bridge between early and late time Universe \cite{6632}, Kantowski-Sachs spherically symmetric solutions \cite{6633}, prospects of constraining $f(T)$ gravity with the third-generation gravitational-wave detectors \cite{6634} and Compact stars in \cite{6635}.
}

There are very few known black hole solutions in any of MTG \cite{24,34,47,51}. The known black hole solutions in MTG are restricted to at most charged massive black holes. Despite the excellent progress in construction of different type of black holes in Einstein gravity, there is no significant progress in construction of black holes in MTG, such as black holes with the Taub-Newman-Unti-Tamburino (Taub-NUT) charges.
Moreover, the rich structure of conformal symmetry in a dual theory to the MTG black holes, is not fully explored and studied. To the best knowledge of the authors, the only article about possible holography for a specific $f(T)$ black hole is the reference \cite{23}. In this article, the authors discovered the existence of the conformal symmetry for an asymptotically anti–de Sitter (AdS) rotating charged black holes in $f(T)$ gravity \cite{47}, where $f(T) = T +\alpha T^2$, where $\alpha$ is a constant. \textcolor{black}{In fact, the idea of AdS/CFT correspondence \cite{11} was extended to the case of extremal rotating black holes, namely, the Kerr/CFT correspondence which was proposed by Guica et al. \cite{22}. The correspondence states that the physics of the extremal Kerr black holes, which are rotating with maximum angular velocity, can be described by a 2D CFT, living on the near-horizon region of the black holes. The correspondence was established by showing that one can microscopically reproduces the Bekenstein-Hawking entropy, using the CFT Card entropy formula. As one would expect, the Kerr/CFT correspondence is not only a peculiar property of extremal black holes but also non-extremal Kerr black holes. However, at the near-horizon region of the non-extremal Kerr black holes, one cannot indicate any conformal symmetries. In other words, the conformal symmetries are not the symmetries of the non-extremal Kerr black hole geometry (as they are for the case of the extremal Kerr black holes). However, it turns out that the “hidden” conformal symmetries can be revealed by looking at the solution space of the radial part of the Klein-Gordon equation, for a massless scalar probe in the near-horizon region of the Kerr black holes \cite{33}. In this case, the radial equation, can be written as the $SL(2, R)_L \times SL(2, R)_R$ Casimir eigen-equation. Subsequently, the Kerr/CFT correspondence can be established by matching the microscopic CFT Cardy entropy to the macroscopic Bekenstein-Hawking entropy of the Kerr black holes with general angular momentum and mass parameters. 
}

Gravitational instantons are the regular and complete solutions to the Einstein field equations. They have self-dual curvature two-form and are asymptotically locally Euclidean \cite{GIBBONS1978430}. 
{\textcolor{black}{In general, the instanton solutions are the result of reduced complex elliptic Monge-Amp\`ere equation on a complex manifold of dimension 2, to only one real variable \cite{666}. The gravitational instantons  play an important role in construction of higher-dimensional solutions to extended theories of gravity \cite{794} and supergravity \cite{888,999}.
}}

There are several well known gravitational instantons, such as Taub-NUT and  Eguchi-Hanson spaces. These solutions play an important role in construction of higher-dimensional extended theories of gravity, supergravity, and the quantum properties of the black holes.
In this article, we construct the Taub-NUT solutions to a specific $f(T)$ gravity and study its behaviours. 
{\textcolor{black}{The article is the first step in constructing the rotating solutions with the NUT twist in the context of $f(T)$ gravity. Once constructed, it is an open question to establish or rule out the existence of the conformal symmetry for the rotating NUT solutions in $f(T)$ gravity. 
}}
The paper is organizes as follows: In section \ref{sec1}, we review the $f(T)$ gravity and the field equations. In section \ref{sec2}, we use a general form for the tetrads in $f(T)$ gravity, and solving all the field equations, we obtain the metric for the Taub-NUT solutions in $f(T)$ gravity. In section \ref{sec3}, we discuss the physical properties of the Taub-NUT solutions in $f(T)$ gravity.

\section{Lorentz covariant $f(T)$ gravity}
\label{sec1}

Einstein's general relativity is a theory built upon the premise of vanishing torsion while being metric compatible and explaining gravity solely with curvature. This makes the fundamental objects the Riemann tensor, Ricci tensor, Ricci scalar and the unique Levi-Civita connection \cite{gravitation}.
TEGR makes use of a relation between the Ricci scalar and the torsion scalar when curvature vanishes, with this one uses the curvatureless Weitzenb{\"o}ck connection\footnote{Some authors claim the Weitzenb{\"o}ck connection to only be the connection that enforces vanishing spin connection. We take the diffrent definition that the Weitzenb{\"o}ck connection is any that has vanishing curvature and non-metricy while keeping torsion.} and uses torsion to describe gravity \cite{tele_book}.

The premise of teleparallel gravity is built upon the fundamental object, the tetrad. This tetrad, $\tensor{h}{^a_\mu}$, is an object that connects the spacetime (Greek index) to the tangent space (beginning of the Latin alphabet index) via
\begin{equation}
    g_{\mu\nu} = \tensor{h}{^a_\mu}\tensor{h}{^b_\nu}\eta_{ab}\:, \label{grel}
\end{equation}
and the inverse relation
\begin{equation}
   \eta_{ab} = \tensor{h}{_a^\mu}\tensor{h}{_b^\nu}g_{\mu\nu}\:. \label{etarel}
\end{equation}
Here $\eta_{ab}$ is the tangent space metric defined to be Minkowski with signature $(+,-,-,-)$ where from (\ref{grel}) and (\ref{etarel}), we find we must have the identities $\tensor{h}{_a^\mu}\tensor{h}{^a_\nu} = \tensor{\delta}{^\mu_\nu},\: \tensor{h}{_a^\mu}\tensor{h}{^b_\mu} = \tensor{\delta}{_a^b}$. The namesake property of TEGR is that the tetrad is always covariantly conserved such that
\begin{equation}
    \partial_\mu \tensor{h}{^a_\nu} + \tensor{\omega}{^a_b_\mu}\tensor{h}{^b_\nu} - \tensor{\Gamma}{^\rho_{\nu\mu}}\tensor{h}{^a_\rho} = 0\:,
\end{equation}
where $\tensor{\Gamma}{^\rho_{\nu\mu}}$ is the aforementioned Weitzenb{\"o}ck connection, and $\tensor{\omega}{^a_b_\mu}$ is the teleparallel spin connection \cite{tele_book}. Thus, the basic pieces are the tetrads $\tensor{h}{^a_\mu}$ and spin connection $\tensor{\omega}{^a_b_\mu}$ of teleparallel theory. With them one builds the torsion tensor as 
\begin{equation}
    \tensor{T}{^a_{\mu\nu}} = \partial_\mu \tensor{h}{^a_\nu} - \partial_\nu \tensor{h}{^a_\mu} + \tensor{\omega}{^a _b _\mu}\tensor{h}{^b_\nu} -\tensor{\omega}{^a _b _\nu}\tensor{h}{^b_\mu}\:,
\end{equation}
and from it the contortion tensor
\begin{equation}
    \tensor{K}{^{\mu\nu}_a} =\dfrac{1}{2}\left(\tensor{T}{_a^{\mu\nu}}+\tensor{T}{^{\nu\mu}_a}-\tensor{T}{^{\mu\nu}_a}\right)\:.
\end{equation}
The contortion tensor allows one to convert from the Levi-Civita connection $\{\tensor{}{^\rho_{\nu\mu}}\} $ to the Weitzenb{\"o}ck connection and vice versa via
\begin{equation}
\tensor{\Gamma}{^\rho_{\nu\mu}} = \{\tensor{}{^\rho_{\nu\mu}}\} +\tensor{K}{^\rho_{\nu\mu}}\:.
\end{equation}
The only unique contraction of the torsion tensor is given by $T^\mu = \tensor{T}{^a ^\mu _a}$, which we can use to make the super potential tensor
\begin{equation}
   \tensor{S}{_a^{\mu\nu}}= \tensor{K}{^{\mu\nu}_a} +T^\mu \tensor{h}{_a^\nu}-T^\nu \tensor{h}{_a^\mu}\:,
\end{equation}
which allows for the definition of the torsion scalar to be $T = \tensor{T}{^a_{\mu\nu}}\tensor{S}{_a^{\mu\nu}}$. This torsion scalar allows the action of $f(T)$ gravity to be written as 
\begin{equation}
    \mathcal{S}_G = \dfrac{1}{4\kappa}\int\dd[4]{x}hf(T)\:,\label{f(T) action}
\end{equation}
where $h=\vert\det(h_{a\mu})\vert$ and\footnote{Note that here and throughout the paper we have set $c=G=1$ for convenience.} $\kappa = 8\pi$. If $f(T)=T$, the $f(T)$ gravity reduces to the teleparallel theory. It can be shown \cite{covariant} that the spin connection is a purely gauge object, and can always be set to zero by choosing a proper tetrad. In fact in the teleparallel theory, the spin connection contributes only a surface term in the Lagrangian, and thus all frames are proper frames \cite{surface_term}. The spin connection obey the transformation law
\begin{equation}
 \tensor{\omega}{^\prime^a_{b\mu}}=\tensor{\Lambda}{^a_c}\tensor{\omega}{^c_{d\mu}}\tensor{\Lambda}{_b^d}+\tensor{\Lambda}{^a_c}\partial_\mu\tensor{\Lambda}{_b^c}\:,
\end{equation}
where $\tensor{\Lambda}{^a_c}$ is a Lorentz matrix and $\tensor{\Lambda}{_a^c}$ is its inverse. For a proper choice of Lorentz matrix, one can make $\tensor{\omega}{^\prime^a_{b\mu}}=0$ and then only consider the proper tetrad $\tensor{h}{^\prime^a_\mu}$ from
\begin{equation}
    \tensor{h}{^\prime^a_\mu} = \tensor{\Lambda}{^a_b}\tensor{h}{^b_\mu}\:.
\end{equation}
This is the importance of Lorentz covariant $f(T)$ theory, transforming from a proper frame introduces spin connection. To properly study $f(T)$ one must make use of proper frames and not just assume $\tensor{\omega}{^\prime^a_{b\mu}}=0$ for any tetrad. Methods to find proper frames range from solving the anti-symmetric field equations \cite{Proper_Frame} or ``turning off" gravity and seeing what inertial effects exist \cite{covariant}. However, as long as you start with and work from a proper frame the final result is guaranteed to be Lorentz covariant. 

With the action (\ref{f(T) action}) one can vary with respect to the tetrad and find the equations of motion to be 
\begin{equation}
    \tensor{\tilde{G}}{_a^\rho}\equiv \dfrac{f(T)}{4}\tensor{h}{_a^\rho} + f_T\left(h^{-1}\partial_\nu\qty(h\tensor{S}{_a^\rho^\nu}) +\tensor{h}{_a^\mu}\tensor{S}{_b^\nu^\rho} \tensor{T}{^b_\nu_\mu} -\tensor{\omega}{^b_a_\mu}\tensor{S}{_b^\rho^\mu}\right) + f_{TT}\tensor{S}{_a^\rho^\mu}\partial_\mu T=\kappa \tensor{\mathcal{T}}{_a^\rho}\:,\label{EOM}
\end{equation}
where  $\tensor{\mathcal{T}}{_a^\rho}$ is the energy-momentum tensor, $f_T=\frac{df}{dT}$ and $f_{TT}=\frac{d^2f}{dT^2}$. The anti-symmetric part of the equation of motion can be shown \cite{antisymm1,antisymm2} to be equivalent to the variation with respect to the spin connection and become
\begin{equation}
    \tensor{\tilde{G}}{_[_a_b_]}\equiv\partial_\mu f_T\qty[\partial_\nu\qty(h\tensor{h}{_{[a}^\mu}\tensor{h}{_{b]}}^\nu)+2h\tensor{h}{_c^{[\mu}}\tensor{h}{_{[a}^{\nu]}}\tensor{\omega}{^c_{b]\nu}}]=0\:.
\end{equation}
Should $f(T)=T$, we find the anti-symmetric part to be trivially satisfied, and that $\tensor{\tilde{G}}{_\mu^\nu}$ reduces to the Einstein tensor.

\section{NUT Solutions in $f(T)$ gravity}
\label{sec2}
When it comes to solving the $f(T)$ field equations, two standard ways exist. One can specify a form of $f(T)$, and use that to find the unknown functions or one can leave it unknown, and solving for it last, after getting the remaining equation into a form containing the torsion scalar. We employ the method of unknown $f(T)$, which gives us an extra degree of freedom in solving the problem. As the Taub-NUT spacetime is an axially symmetric spacetime \cite{Taub,NUT,exact}, we employ the tetrad 
\begin{equation}
    {h^a}_\mu = \begin{pmatrix}
    {A(r)} & 0 & 0 & {A(r)}(c_1\cos\theta+c_2) \\
    0 & {B(r)}\sin\theta\cos\varphi & {C(r)}\cos\theta\cos\varphi & -{C(r)}\sin\theta\sin\varphi \\
    0 & {B(r)}\sin\theta\sin\varphi & {C(r)}\cos\theta\sin\varphi  & {C(r)}\sin\theta\cos\varphi \\
    0 & {B(r)}\cos\theta & -{C(r)}\sin\theta  &0
    \end{pmatrix}\:.\label{gen_tet}
\end{equation}
The tetrad (\ref{gen_tet}) solves the anti-symmetric field equations for any functions $A(r),B(r),C(r)$ and constants $c_1$ and $c_2$ \cite{Proper_Frame}. This leaves us with 4 unknown functions, and 2 unknown constants. Through (\ref{grel}) with tetrad (\ref{gen_tet}), we find the metric to take the form
\begin{equation}
    ds^2 = A^2(r)dt^2-B^2(r)dr^2-C^2(r)d\theta^2 -(C^2(r)\sin^2\theta - A^2(r)(c_1\cos\theta+c_2)^2)d\varphi^2+2A^2(r)(c_1\cos\theta+c_2)dtd\varphi\:.
\end{equation}
One can see that with following substitutions, we find the traditional Taub-NUT metric
\begin{equation}
    \begin{matrix}
        A(r)=\dfrac{1}{B(r)} = \sqrt{\dfrac{r^2-2mr-n^2}{r^2+n^2}}\:, && C(r)=\sqrt{r^2+n^2}\:, && c_1 = 2n\:, && c_2 = 0\:.\label{Taub-NUT-Standard}
    \end{matrix}
\end{equation}
We first make the substitution for $c_1$ and $c_2$ from (\ref{Taub-NUT-Standard}) to simplify the tetrad (\ref{gen_tet}) to the form
\begin{equation}
    {h^a}_\mu = \begin{pmatrix}
    {A(r)} & 0 & 0 & 2nA(r)\cos\theta \\
    0 & {B(r)}\sin\theta\cos\varphi & {C(r)}\cos\theta\cos\varphi & -{C(r)}\sin\theta\sin\varphi \\
    0 & {B(r)}\sin\theta\sin\varphi & {C(r)}\cos\theta\sin\varphi  & {C(r)}\sin\theta\cos\varphi \\
    0 & {B(r)}\cos\theta & -{C(r)}\sin\theta  &0
    \end{pmatrix}\:.\label{beg_tet}
\end{equation}
With tetrad (\ref{beg_tet}), we may compute the field equations from (\ref{EOM}) subject to $ \tensor{\mathcal{T}}{_\mu_\nu} = 0$. Doing so and simplifying as much, we find the following field equations:  temporal component
\begin{align}
\tensor{\tilde{G}}{^t_t} &= \dfrac{f(T)}{4} + \dfrac{2f_T}{AC^4B^3}\qty(B^3A^3n^2+C^3B^2A'+C'\qty[AC^2\qty(B^2+B'C)-BA'C^3]-C^3BC''-C^2(C')^2BA)\nonumber\\&-\dfrac{2f_{TT}T'}{B^2C}(C' -B)\:,
\end{align}
radial component
\begin{equation}
\tensor{\tilde{G}}{^r_r} = \dfrac{f(T)}{4} + \dfrac{2f_T}{AB^2C^2}\qty(C'AB+CBA'-(C')^2A-2CC'A')\:,
\end{equation}
angular components
\begin{align}
\tensor{\tilde{G}}{^\theta_\theta} &=\tensor{\tilde{G}}{^\varphi_\varphi}=\dfrac{f(T)}{4} + \dfrac{f_T}{AC^4B^3}\left(A'\qty[2B^2C^3+C^4B']+C'\qty[2AB^2C^2+AC^3B'-3BA'C^3]-2B^3A^3n^2\right.\nonumber\\
&-\left.C^2B^3A-(C')^2C^2BA-C^3BAC''-C^4BA''\right)-\dfrac{2f_{TT}T'}{AB^2C}(A(B-C')-A' C)\:,
\end{align}
and rotational component
\begin{align*}
\tensor{\tilde{G}}{^t_\varphi} &= \dfrac{2n\cos\theta f_T}{AC^4B^3}\qty(A''C^4B-C^3BAC''-(C')^2C^2BA+C^3C'\qty[B'A+BA']-C^4B'A'+AB^3\qty[4n^2A^2+C^2])\\
&-\dfrac{2nf_{TT}T'\cos\theta}{AB^2C}(A(B-C')+A' C)\:.\yesnumber
\end{align*}
In addition to the field equations, we also find the torsion scalar to take the form \footnote{Where for shorthand we defined $A(r)\equiv A$ and $A'$ represents a derivative with respect to $r$.}
\begin{equation}
    T = \dfrac{4}{AB^2C^4}\qty(n^2A^3B^2+AC^2(C')^2-2C'ABC^2+2C'A'C^3+AB^2C^2-2BA'C^3)\:.\label{TS}
\end{equation}
It can be shown \cite{Biancci_identity} that for $\tensor{\imaginary}{^\mu_\nu} \equiv \tensor{\tilde{G}}{^\mu_\nu}-\kappa\tensor{\mathcal{T}}{^\mu_\nu}=0$, should the anti-symmetric equations be satisfied, we have the identity that
\begin{equation}
    \mathcal{D}_\mu\tensor{\imaginary}{^\mu_\nu} = 0\:,\label{DG}
\end{equation}
where $\mathcal{D}_\mu$ is the Levi-Civita covariant derivative. This tells us as long as $\tensor{\imaginary}{^r_r}$ is satisfied, we only need satisfy either the temporal or angular equation, and the other is automatically satisfied. This is seen via the $\nu=r$ component of (\ref{DG})
\begin{equation}
    \mathcal{D}_\mu\tensor{\imaginary}{^\mu_r} = \partial_r \tensor{\imaginary}{^r_r} +\qty(\dfrac{A'}{A} + \dfrac{2C'}{C})\tensor{\imaginary}{^r_r} -\dfrac{A'}{A}\tensor{\imaginary}{^t_t}+\dfrac{C'}{C}\qty(\tensor{\imaginary}{^\theta_\theta}+\tensor{\imaginary}{^\varphi_\varphi})=0.
\end{equation}
With this in mind,  we follow an approach for spherical symmetric problems \cite{Spherical} and use the angular equation to solve for the term $f_{TT}T'$, to eliminate it from both the temporal and rotational equations. We reserve the radial equation for the use of solving for the form of $f(T)$, but use it to eliminate the $f(T)$ dependence in the temporal and rotational equations in favour of $f_T$. After such work, we find the field equations to take the forms
\begin{align*}
    \tensor{\tilde{G}}{^t_t} &= \frac{2f_T}{B^{2}C^{4}\left(\left(B-C'\right)A-A'C\right)A}\left(3A^{4}B^{3}n^{2}-3A^{4}B^{2}C'n^{2}-A^{3}B^{3}A'Cn^{2}+AC^{4}A''B-AC^{4}A''C'\right.\\&+\left.AC^{4}C''A'-B'AA'C^{4}+A^{2}B^{3}C^{2}-A^{2}B^{2}C'C^{2}-A^{2}B\qty(C')^{2}C^{2}+A^{2}\left(C'\right)^{3}C^{2}-C^{4}\left(A'\right)^{2}C'\right)\yesnumber\:,
\end{align*}
\begin{align*}
\tensor{\tilde{G}}{^t_\varphi}&= \frac{4\cos(\theta)f_Tn}{A\left(A\left(B-C'\right)-A'C\right)C^{4}B^{2}}\left(3A^{4}B^{3}n^{2}-3A^{4}B^{2}C'n^{2}-A^{3}B^{3}A'Cn^{2}+AC^{4}A''B-AC^{4}A''C'\right.\\&+\left.AC^{4}C''A'-B'AA'C^{4}+A^{2}B^{3}C^{2}-A^{2}B^{2}C'C^{2}-A^{2}B\qty(C')^{2}C^{2}+A^{2}\left(C'\right)^{3}C^{2}-C^{4}\left(A'\right)^{2}C'\right)\yesnumber\:,
\end{align*}
\begin{equation}
\tensor{\tilde{G}}{^r_r} = \frac{f(T)AB^2C^2+f_T\qty(8ABC'-8A\left(C'\right)^2+8BCA'-16CC'A')}{4AB^2C^2}\:.
\end{equation}
It's interesting to note that the non-trivial temporal and rotational equations become same for the case of vanishing $T_{\mu\nu}$. In fact we have $\tensor{\tilde{G}}{^t_\varphi}=2n\cos\theta \tensor{\tilde{G}}{^t_t}$, so we only need look at one or the other.
This leaves us with 2 equations for 4 unknown functions, from which we cannot generate a solution without more information or assumptions. The first natural assumption is to let $C(r)$ take the form as in (\ref{Taub-NUT-Standard}), while also assume that $A(r) = 1/B(r)$ still holds. These two assumptions make the field equations take the form
\begin{align}
    \tensor{\tilde{G}}{^t_t} &=\dfrac{2f_T\qty(r\sqrt{r^2+n^2}A-r^2-n^2)}{(n^2+r^2)^{5/2}\qty((n^2+r^2)A'+Ar-\sqrt{r^2+n^2})}\left(A(r^2+n^2)^2A''+(r^2+n^2)^2(A')^2+(3n^2-r^2)A^2+n^2+r^2\right)\:,\label{ttfinal}\\
    \tensor{\tilde{G}}{^r_r}&=\dfrac{\sqrt{r^2+n^2}\qty[f(T)(r^2+n^2)^2-8rf_TA\qty(rA+2(r^2+n^2)A')]+8f_T(r^2+n^2)(A'(r^2+n^2)+Ar)}{4(r^2+n^2)^{5/2}}\:\label{rrfinal}.
\end{align}
Solving the temporal and rotational equations, we find two unique solutions. One can easily solve the analytic equation in (\ref{ttfinal}), for $A(r)$ to find
\begin{align}
    A(r) = \dfrac{\sqrt{r^2+n^2}}{r}\:,\label{Sqrtsol}
\end{align}
or the differential equation\footnote{This differential equation is actually a linear differential equation using the substitution $A^2(r)\rightarrow U(r)$} in (\ref{ttfinal}), to find
\begin{equation}
    A(r) = \sqrt{\dfrac{r^2-2C_1r-n^2+C_2(3n^4-6n^2r^2-r^4)}{(r^2+n^2)}}\:.\label{TNsol}
\end{equation}
We note through dimensional analysis that the constants $C_1$ and $C_2$ are related to the mass and cosmological constant, respectively. The latter is better seen if (\ref{TNsol}) is plugged into the radial equation (\ref{rrfinal}), and solved for $f(T)$. We find the functional form $f(T) = T-12C_2$. With (\ref{TNsol}) \textcolor{black}{describing} the standard cosmological $\Lambda$-Taub-NUT solution in Einstein gravity, we turn our attention to (\ref{Sqrtsol}) with (\ref{TS}) to find 
\begin{equation}
    T = \dfrac{4n^2}{r^2\qty(r^2+n^2)}\:.
\end{equation}
The radial field equation (\ref{rrfinal}) with (\ref{Sqrtsol}),  leads to 
the differential equation for $f(T)$ function
\begin{equation}
    f(T) = -2\qty(\dfrac{4n^2}{r^2\qty(r^2+n^2)})\dfrac{df}{dT} = -2T\dfrac{df}{dT}\:.\label{f(T) diff eq}
\end{equation}
We solve (\ref{f(T) diff eq}) and find
\begin{equation}
    f(T) = \dfrac{C}{\sqrt{T}}\:.\label{fTinv}
\end{equation}
for any constant $C$. \textcolor{black}{We can absorb the constant $\kappa$ into $C$} and define the gravitational action (\ref{f(T) action}) to be 
\begin{equation}
\textcolor{black}{\mathcal{S}_G = \dfrac{C}{4}\int\dd[4]{x}\dfrac{h}{\sqrt{T}}\:,}\label{actionfin}
\end{equation}
while the metric takes the form
\begin{equation}
    g_{\mu\nu} = \begin{pmatrix}
        \dfrac{r^2+n^2}{r^2} &0 &0&\dfrac{2n(n^2+r^2)\cos\theta}{r^2}\\
        0 &-\dfrac{r^2}{r^2+n^2}&0&0\\
        0&0&-(r^2+n^2)&0\\
        \dfrac{2n(n^2+r^2)\cos\theta}{r^2}&0&0&\dfrac{r^2+n^2}{r^2}\qty(4n^2\cos^2\theta-r^2\sin^2\theta)
    \end{pmatrix}\:,
\end{equation}
with the tetrad
\begin{equation}
    {h^a}_\mu = \begin{pmatrix}
    \sqrt{r^2+n^2}/r & 0 & 0 & 2n\qty(\sqrt{r^2+n^2}/r)\cos\theta \\
    0 & \qty(r/\sqrt{r^2+n^2})\sin\theta\cos\varphi & \sqrt{r^2+n^2}\cos\theta\cos\varphi & -\sqrt{r^2+n^2}\sin\theta\sin\varphi \\
    0 & \qty(r/\sqrt{r^2+n^2})\sin\theta\sin\varphi & \sqrt{r^2+n^2}\cos\theta\sin\varphi  & \sqrt{r^2+n^2}\sin\theta\cos\varphi \\
    0 & \qty(r/\sqrt{r^2+n^2})\cos\theta & -\sqrt{r^2+n^2}\sin\theta  &0
    \end{pmatrix}\:.\label{solsqrt}
\end{equation}
With this solution, one can set the functions $B(r),C(r)$ and $f(T)$ as defined above, while leaving $A(r)$ unknown. Doing so one finds the simple differential equation to satisfy the system
\begin{equation}
    (n^2 + r^2)^2A' + n^2rA^3 = 0\:,
\end{equation}
whose solution allows the generalization
\begin{equation}
    {h^a}_\mu = \begin{pmatrix}
    \sqrt{\frac{(r^2+n^2)}{\epsilon(r^2+n^2)-n^2}} & 0 & 0 & 2n\qty(\sqrt{\frac{(r^2+n^2)}{\epsilon(r^2+n^2)-n^2}})\cos\theta \\
    0 & \qty(r/\sqrt{r^2+n^2})\sin\theta\cos\varphi & \sqrt{r^2+n^2}\cos\theta\cos\varphi & -\sqrt{r^2+n^2}\sin\theta\sin\varphi \\
    0 & \qty(r/\sqrt{r^2+n^2})\sin\theta\sin\varphi & \sqrt{r^2+n^2}\cos\theta\sin\varphi  & \sqrt{r^2+n^2}\sin\theta\cos\varphi \\
    0 & \qty(r/\sqrt{r^2+n^2})\cos\theta & -\sqrt{r^2+n^2}\sin\theta  &0
    \end{pmatrix}\:.\label{sqrtT}
\end{equation}
We note that should $\epsilon = 1$, (\ref{sqrtT}) reduces to tetrad (\ref{solsqrt}). With this generalization, we find the torsion scalar to become
\begin{equation}
    T = \dfrac{4n^2}{\qty(\epsilon(r^2+n^2)-n^2)\qty(r^2+n^2)}\:.
\end{equation}
\textcolor{black}{Most applicable functional forms in $f(T)$ theory take the form of $f(T) = T + F(T)$, and make use of the FLRW metric $g_{\mu\nu} = \diag(1,a(t),a(t),a(t))$ in which $F(T)$ (or $f(T)$) sometimes contains inverse powers, such as $f(T)\sim \frac{1}{T^{0.11}}$ \cite{ft_cos_data,ft_cos}, like ours (\ref{fTinv}) . Solutions solving for spacetimes outside of these cosmological metrics tend to be much more complex with multiple metric functions and non zero metric components. In these cases improper tetrads tend to be employed or trivial solutions found.\footnote{\textcolor{black}{Trivial solutions in this case refers to solutions where the torsion scalar is constant or solutions for unknown $f(T)$.}} Cases with static spherically symmetric (SSS) spacetimes have a lot of solutions from proper to improper tetrads \cite{wang,Daouda,Noether,prop}. Our solution sits outside of these cases, while still managing to have a proper non-trivial solution. While it may appear the functional form does not coincide with standard forms from cosmology, we can't rule out that the found functional form has no use here or in other areas of physics, for example, naked singularity physics.}
\section{Physical properties}
\label{sec3}
In computation of the physical properties of the newfound spacetime, we analyze the singularities, test particle orbits and asymptotic conditions. Beginning with singularity analysis, we compute the Ricci and Kretschmann scalar invariants. These are found to take the forms
\begin{equation}
    R = \textcolor{black}{\dfrac{6n^4}{(n^2+r^2)(\epsilon(r^2+n^2)-n^2)^2}}\:,\label{Ricci}
\end{equation}
and
\begin{equation}
    \mathcal{K}= \textcolor{black}{\dfrac{-12n^2\qty[(n^2 + r^2)^3\epsilon^3-10n^2(6n^4 +2n^2r^2 +r^4)\epsilon^2+6n^4(n^2 +r^2)\epsilon-3n^6]}{(n^2+r^2)^2(\epsilon(r^2+n^2)-n^2)^4}}\:.\label{Kretschmann}
\end{equation}
From both we note a non-physical imaginary singularity at $r_s=\pm in$. The other singularity will be denoted by $r_s$, and is located at $r=\pm n\sqrt{\frac{1-\epsilon}{\epsilon}}$. For many values of $\epsilon$, {\textcolor{black} {where $\epsilon < 0$ or $\epsilon >1$, this is found to also be a un-physical imaginary singularity, so our solution is regular everywhere, without any type of naked singularities. Analysis of orbits in section \ref{sec3} leads one to see the $\epsilon < 0$ region is un-physical.} It is only the range $0<\epsilon\leq1$ where $r_s$ becomes a spherical surface of naked singularities. Letting $\epsilon=1$, we find the surface to shrink to the single point $r=0$. The existence of naked singularities in our universe, is still an ongoing debate with many assuming the cosmic censorship hypothesis \textcolor{black}{(CCH)} \cite{Penrose} to hold while others think they might be a quantum gravity problem \cite{naked}. Whatever the case may be, this spacetime has one that can be scaled by the value of $\epsilon$\textcolor{black}{, what we can't say is that this disproves the CCH as it only holds for GR solutions. We can only say and show with this found example that naked singularities can arise more often in modified theories. Based off the value of $\epsilon$ it} allows us to find the radial coordinate range to be in one of two options given by $r\in (r_s,\infty)$ for a singularity or $r\in (-\infty,\infty)$ for none. In addition to the radial singularity, one finds Misner strings in the same way that the standard Taub-NUT has, from the line element
\begin{align}
    ds^2 &= \frac{(r^2+n^2)}{\epsilon(r^2+n^2)-n^2}dt^2-\dfrac{r^2}{r^2+n^2}dr^2-(r^2+n^2)d\theta^2 -\qty((r^2+n^2)\sin^2\theta - \frac{4n^2(r^2+n^2)\cos^2\theta}{\epsilon(r^2+n^2)-n^2})d\varphi^2\nonumber\\&+\frac{4n(r^2+n^2)\cos\theta}{\epsilon(r^2+n^2)-n^2}dtd\varphi\:,\label{line ele}
\end{align}
containing $\cos\theta$. 
One may assume the value $\epsilon=0$ may simplify and remove some complexity, as in this case we find $r_s\rightarrow\infty$. However, for the limit that $r$ gets very large, we find the metric becomes
\begin{equation}
    g_{\mu\nu} \approx \begin{pmatrix}
        1/\epsilon &0 &0&2n\cos\theta/\epsilon\\
        0 &-1&0&0\\
        0&0&-r^2&0\\
        2n\cos\theta/\epsilon&0&0&-r^2\sin^2\theta
    \end{pmatrix}\:.\label{reduced met}
\end{equation}
From (\ref{reduced met}), we note that time coordinate goes to infinity for $\epsilon=0$, so this value should be excluded from allowed $\epsilon$ values. From the radial singularity and the angular Misner strings singularities, we can conclude that the metric (\ref{line ele}) is locally asymptotically flat, just as the standard Taub-NUT is \cite{exact}. Under the limit that $n\rightarrow0$ one finds the metric to reduce to the familiar Minkowski space with the $\epsilon$ parameter
\begin{equation}
    \lim_{n\rightarrow 0}g_{\mu\nu} \rightarrow\begin{pmatrix}
        1/\epsilon &0 &0&0\\
        0 &-1&0&0\\
        0&0&-r^2&0\\
        0&0&0&-r^2\sin^2\theta
    \end{pmatrix}\:,\label{reduced met n}
\end{equation}
with $T=0$. This limit is however problematic as our functional form for $f(T)$ diverges for $T=0$ and thus our action (\ref{actionfin}) diverges too. \textcolor{black}{This leads one to take 2 different actions. We can take this limit before the solution of (\ref{ttfinal}) in which we find the equation} 
\begin{equation}
    \tensor{\tilde{G}}{^t_t} =\dfrac{2f_T\qty(A-1)}{r^2\qty(rA'+A-1)}\left(r^2AA''+r^2(A')^2-A^2+1\right)\:.\label{ttlimit}
\end{equation}
The solutions take the form $A=1$ or the de Sitter–Schwarzschild metric. With $A=1$, we find the radial equation reduces to $f(T)=0$, which is standard as $T=0$ describes Minkowski space. \textcolor{black}{Separately we may make the  mandate}
\begin{equation}
    \textcolor{black}{\lim_{n\rightarrow0}\dfrac{C}{\sqrt{T}}\equiv 0}\:.\label{action n lim}
\end{equation}
\textcolor{black}{This prevents the problem in the action as it circumvents the $T=0$ problem. Enforcing this constraint makes the constant $C$ take the form $C=n^{2}\tilde{C}$ while $\tilde{C}$ is a new constant. We then have no further constraints on $\tilde{C}$ aside from it must finite for $n=0$. Hence, our solution clearly has a Minkowski limit.}
Should we turn our attention to orbits, we find from (\ref{line ele}) that we have no dependence on $t$ and $\varphi$, as there are 2 Killing vectors. We assume $\theta = \pi/2$, and denote $p_t$ and $p_\varphi$ as constants of motion. Analysis of the constants, we find $p_t$ is related to the energy of the particle, which an observer measures at infinity 
\begin{equation}
    \eval{E}_{r=\infty} = p_t\epsilon\:,
\end{equation}
while $p_\varphi$ is related to the angular momentum via
\begin{equation}
    p_t = \eval{E}_{r=\infty} v_{\hat{\varphi}} \sqrt{r^2+n^2}\:,
\end{equation}
where $v_{\hat{\varphi}}$ is the tangential velocity, measured by an observer in an equatorial trajectory. With this in mind, we may compute the radial orbit equation to become
\begin{equation}
    g^{\sigma\nu}p_\sigma p_\nu = -\mu^2 \Rightarrow\qty(\dfrac{dr}{d\lambda})^2 = \dfrac{C_1}{r^2}+C_2\:,\label{reqn}
\end{equation}
with the constants $C_1$ and $C_2$ defined as 
\begin{equation}
    C_1 = n^2\mu^2+n^2p_t^2(\epsilon-1)-p_\varphi^2\:,\label{c1}
\end{equation}
and
\begin{equation}
    C_2 =\epsilon p_t^2+\mu^2\:,\label{c2}
\end{equation}
with $\mu$ being the test particles mass. With this equation it can be shown that no stable orbits can be created in this spacetime. What is to be said for this spacetime, is that all orbit equations have closed form analytical solutions. The solution to (\ref{reqn}) is found to be  
\begin{equation}
    r(\lambda) = \sqrt{C_2(\beta+\lambda)^2-\dfrac{C_1}{C_2}}\:,\label{r sol}
\end{equation}
where $\lambda$, is the affine parameter defined for massless particles, and equal to $\tau/\mu$ for massive ones, and $\beta$ is an arbitrary constant. With a solution found for $r(\lambda)$, both $t(\lambda)$ and $\varphi(\lambda)$ can be found. Computing their differential equations, we find
\begin{equation}
    \dv{\varphi}{\lambda} =-\dfrac{p_\varphi}{n^2+r^2}= -\dfrac{p_\varphi}{C_3+C_2(\beta+\lambda)^2}\:,
\end{equation}
and 
\begin{equation}
    \dv{t}{\lambda} = -p_t\dfrac{\epsilon(n^2+r^2)-n^2}{n^2+r^2}=p_t\epsilon - \dfrac{n^2p_t}{C_3+C_2(\beta+\lambda)^2}\:.
\end{equation}
Due to the similar form of the differential equations, it's no surprise the solutions come out to be very similar in form with 
\begin{equation}
    \varphi(\lambda) =-\dfrac{p_\varphi C_2}{\sqrt{C_2^3n^2-C_1C_2^2}}\tan[-1](\dfrac{C_2^2(\beta + \lambda)}{\sqrt{C_2^3n^2-C_1C_2^2}})-\alpha\:.\label{varphi sol}
\end{equation}
and 
\begin{equation}
    t(\lambda) = p_t\epsilon\lambda -\dfrac{p_tn^2C_2}{\sqrt{C_2^3n^2-C_1C_2^2}}\tan[-1](\dfrac{C_2^2(\beta+\lambda)}{\sqrt{C_2^3n^2-C_1C_2^2}}) + \delta\:,
\end{equation}
where $\alpha$ and $\delta$, are again arbitrary constants. What is typically more important is $r(\varphi)$, showing the total angle swept out in the trajectory. To do so, we can solve (\ref{varphi sol}) for $\lambda$ to obtain $\lambda(\varphi)$ which becomes
\begin{equation}
    \lambda(\varphi) = -\beta - \dfrac{\chi p_\varphi}{\textcolor{black}{C_2}}\tan(\chi(\varphi+\alpha))\:,\label{lambda}
\end{equation}
with $\chi$, a new constant, defined as
\begin{equation}
    \chi \equiv \dfrac{\sqrt{\textcolor{black}{C_2}^2(n^2\textcolor{black}{C_2}-\textcolor{black}{C_1})}}{p_\varphi \textcolor{black}{C_2}}\:.
\end{equation}
Substitution of (\ref{lambda}) into (\ref{r sol}), we find
\begin{equation}
    r(\varphi) = \sqrt{\dfrac{\sin[2](\chi (\varphi+\alpha))n^2\textcolor{black}{C_2}-\textcolor{black}{C_1}}{\textcolor{black}{C_2}\cos[2](\chi(\varphi+\alpha))}}\:.
\end{equation} 
Looking at trajectory's, we find that the singularity acts as a photon mirror, with photons regardless of energy cannot hit the singularity for $\epsilon$ in $0<\epsilon\leq1$ as long as $p_\varphi\neq 0$. \textcolor{black}{This photon mirror property gives rise to a clear astrophysical test to distinguish this/any photon mirror spacetime from standard Taub-NUT/non-photon mirror spacetimes. We could search the sky for dark naked singularities that when found reflect all light towards them.} For massive particles, we find that they do hit the singularity for a certain value of $\lambda$. This can be found by solving (\ref{r sol}) for $\lambda$, when it equals the location of the singularity, to find
\begin{equation}
    r(\lambda) = n\sqrt{\dfrac{1-\epsilon}{\epsilon}} \Rightarrow \lambda =\pm\sqrt{\dfrac{m^2n^2-p_\varphi^2\epsilon}{\epsilon(m^2+\epsilon p_t^2)^2}}-\beta\:.
\end{equation}
When $m=0$ we note that the square root always is imaginary, unless $p_\varphi=0$ showing no photon orbits cross the singularity unless sent with no angular momentum. Massive particles will hit the singularity if $m^2n^2>\epsilon p_\varphi^2$. Once $\epsilon\geq 1$ we have no singularity, however, orbits still acts strange. All orbits are still pulled into the origin if $c_1\geq0$, or reflected if $c_1<0$ for both massive and massless particles, and we never see orbits go around the origin. When $\epsilon<0$ we have the added condition that $m^2>-\epsilon p_t^2$ for trajectory's to exist, telling us no photon trajectory's exist in negative $\epsilon$ spacetimes. Furthermore for massive particles the condition for $c_1$ remains the same  and all trajectory's exhibit reflective nature from the origin. Further research is needed to be undertaken to know if this is due to $f(T)$ theory, the inverse nature of torsion in the functional $f(T)$ form, or a mixed property of \textcolor{black}{whichever} and the NUT charge. \textcolor{black}{We suspect the photon mirror property to be related to the inverse nature of $f(T)$ form and thus some singularity properties can arise form the functional form, but we cannot say for certain without further research. Plans exist to better generalize the existing $f(T)$ form to look more like standard forms.}

\section{Conclusion}
\textcolor{black}{Gravitational instantons are one of the most interesting class of exact solutions in general relativity. The gravitational instantons  are regular and make complete solutions  everywhere. One special feature of gravitational instantons is existence of self dual curvature two form.  The gravitational instantons exist in vacuum \cite{791}, as well as in presence of the cosmological constant \cite{792}. One important class of gravitational instantons is Taub-NUT solution.
The Taub-NUT solution and its extensions have been used extensively in higher-dimensional gravity \cite{794,7941}, M-theory \cite{795,7951}, black hole holography \cite{796} and black hole thermodynamics \cite{797}. Despite the existence and success of gravitational instantons in general relativity, there are no known such solutions in $f(T)$ gravity.
}

Inspired by the existence of self-dual geometries in Einstein gravity, we construct exact analytical solutions to the $f(T)$ gravity theory, resembling the Taub-NUT solutions.  In this regard, we consider the field equations of $f(T)$ gravity with an unknown $f(T)$ function. 
{\textcolor{black}{We start with considering a well defined proper tetrad basis which is in agreement with the fact that Taub-NUT spacetime is an axially symmetric spacetime. After that, we build the field equations of $f(T)$ gravity with no energy-momentum tensor.
We solve analytically, without any approximations, all the field equations to find the exact solutions for the Taub-NUT spacetime, as well as the functional form of the $f(T)$ function.}} To our knowledge, this is the first construction of the Taub-NUT solutions in $f(T)$ gravity {\textcolor{black}{with the proper tetrad}} and non-zero torsion. We discuss the physical properties of the solution. {\textcolor{black}{Especially, by calculating the Ricci scalar and the Kretschmann scalar, we find that for a large amount of parameter $\epsilon$, the solutions are regular everywhere, free of any kind of singularities. The solutions at the limit of $n \rightarrow 0$ approaches to the Minkowski spacetime. We also discuss the behaviour of orbits of a test particle.}}

{\textcolor{black}{The cosmological evolution in $f(T)$ gravity is considered mainly with other functions for $f(T)$ gravity. As an example, researchers have extensively explored power-law models of $f(T)=T+F(T)$ gravity to reconstruct and describe various evolution scenarios of the universe \cite{P1}, where $F(T)$ (or $f(T)$) sometimes contains inverse powers, such as $f(T) \sim \frac{1}{T^{0.11}}$ \cite{ft_cos_data,ft_cos}.  Finding the analytical solutions for the spacetime tend to be much more complex with multiple metric functions and non zero metric components. In all these cases improper tetrads tend to be employed. Even cases such as SSS  spacetimes have a lot of solutions from proper to improper tetrads \cite{wang,Daouda,Noether,prop}. Our solution in this article, is outside of these cases, which still manages to have a proper non-trivial solution based on a proper tetrad (\ref{sqrtT}). While it may appear the functional form does not coincide with standard forms from cosmology, we can’t rule out that the found functional form has no use here or in other areas of physics, for expample, naked singularity physics. So although it is not feasible to verify the compatibility of the result of the present article with $f(T)=C/\sqrt{T}$, with their results, we plan to continue our research in a forthcoming article to discuss the cosmological evolution with  $f(T)=C/\sqrt{T}$, which allows us to compare the results of the present article with cosmological evolution. We also mention that power law $f(T)$ gravity has been employed successfully  in the context of the linear growth in power law $f(T)$ gravity \cite{P2}, cosmological viscous fluid models describing infinite time singularities in $f(T)$ gravity \cite{P3},  astrophysical observations compatibility with $f(T)$ gravity \cite{P4}, resolving FLRW cosmology through effective equations of state in $f(T)$ gravity with power law \cite{P5}, exploring late-time cosmic acceleration \cite{P6}, prospects of constraining $f(T)$ gravity with the third-generation gravitational-wave detectors \cite{P7}, cosmic growth in $f(T)$ teleparallel gravity \cite{P8}.}}

{\textcolor{black}{Moreover the cosmological models within $f(T)$ gravity were considered in the following articles: attractor behaviour of $f(T)$ modified gravity and the cosmic acceleration \cite{P9}, $f(T)$  cosmology from pseudo-bang to pseudo-rip \cite{P10}, cosmic evolution in $f(T)$ gravity theory \cite{P11}, anisotropic cosmological dynamics in $f(T)$ gravity in the presence of a perfect fluid \cite{P12}, type IV singular bouncing cosmology from $f(T)$ Gravity \cite{P13}, interacting dark energy in $f(T)$ cosmology \cite{P14}, locally rotationally symmetric Bianchi type-I cosmological model in f(T) gravity \cite{P15}.
}}

{\textcolor{black}{For future works, we plan to construct the analog of other self-dual geometries, such as Eguchi-Hanson in $f(T)$ gravity, and discuss their physical properties.
}}

{\textcolor{black}{Moreover, in the context of the duality between the rotating black holes and the conformal field theory (CFT) in general relativity, we may find and look at the rotating solutions to the results of this article, to establish the existence (or non existence) of such duality for the rotating black hole solutions in $f(T)$ gravity. The duality has been shown to be valid through comparison between the macroscopic black hole quantities, as solutions of the general relativity, and the microscopic CFT quantities. In particular, in the context of duality, there is a perfect match between the macroscopic Bekenstein-Hawking entropy of the rotating black holes and the entropy of the CFT which is computed by the Cardy formula. Another very interesting result which supports the duality is coming from the study of the super-radiant scattering off the rotating black holes. It was shown that the bulk scattering amplitudes are in precise agreement with the scattering results from CFT. The scattering amplitudes of CFT are completely by the conformal invariance. 
}

\section*{Acknowledgments}
The authors would like to thank Bardia H. Fahim for discussions. This work was supported by the Natural Sciences and Engineering Research Council of Canada.
\section*{References}
\bibliographystyle{unsrt}
\bibliography{refs}
\end{document}